\documentclass{article}

\usepackage{arxiv}

\usepackage[utf8]{inputenc} 
\usepackage[T1]{fontenc}    
\usepackage{hyperref}       
\usepackage{url}            
\usepackage{booktabs}       
\usepackage{amsfonts}       
\usepackage{nicefrac}       
\usepackage{microtype}      
\usepackage{lipsum}		
\usepackage{graphicx}

\usepackage{amsmath}
\usepackage{algorithm}
\usepackage{algorithmicx}
\usepackage{algpseudocode}
\usepackage{subfigure}

\title{Encoding Candlesticks as Images for Pattern Classification Using Convolutional Neural Networks}

\author{Jun-Hao Chen\\
	Department of Computer Science \& Information Engineering\\
	National Taiwan University\\
	Taipei 10617, Taiwan\\
	\And
	Yun-Cheng Tsai\\
	School of Big Data Management\\
	Soochow University\\
	Taipei 11102, Taiwan\\
	\texttt{pecutsai@gm.scu.edu.tw}
}

\begin{document}
\maketitle

\begin{abstract}
Candlestick charts display the high, low, opening, and closing prices in a specific period. Candlestick patterns emerge because human actions and reactions are patterned and continuously replicate. These patterns capture information on the candles. According to Thomas Bulkowski's Encyclopedia of Candlestick Charts, there are $103$ candlestick patterns. Traders use these patterns to determine when to enter and exit. Candlestick pattern classification approaches take the hard work out of visually identifying these patterns. To highlight its capabilities, we propose a two-steps approach to recognize candlestick patterns automatically. The first step uses the Gramian Angular Field (GAF) to encode the time series as different types of images. The second step uses the Convolutional Neural Network (CNN) with the GAF images to learn eight critical kinds of candlestick patterns. In this paper, we call the approach GAF-CNN. In the experiments, our approach can identify the eight types of candlestick patterns with $90.7\%$ average accuracy automatically in real-world data, outperforming the LSTM model.
\end{abstract}


\section{Introduction}
\label{S:1}
Financial market forecasts are critical research topics in commercial finance and information engineering. For example, the topics are predicting fluctuations or volatility forecasts for futures indices~\cite{kou2014evaluation}. Market prices are susceptible to the expected psychological impact of the overall market. These prices are possible to develop predictive models of financial demand through particular pre-processing and complex model architectures.

Many tools are existing to help people predict stock price fluctuations and futures indices already~\cite{ding2015deep}. For example,  these tools are the neural networks, fuzzy time-series analysis, genetic algorithms, classification trees, statistical regression models, and support vector machines. However, these machine learning models are generic techniques and used for forecasting. They unusually combine with financial expertise~\cite{kou2014evaluation}. Because the average person pursues profit in any transaction, the predictions of such models are not accurate enough for real-world operations. Investment forecasts and model predictions tend to have significant gaps, and investors are more inclined to find a good entry and exit point rather than merely predicting prices.  Many studies focus on the accuracy of numerical predictions~\cite{saad1998comparative, refenes2001forecasting, pantazopoulos1998financial, dhar2001comparison, cao2003support, song1993fuzzy}, but investors only concern with the time of entry and exit (i.e., how much profit space they have). In other words, rather than blindly using machine learning or deep learning architecture to pursue unrealistic low-risk, high-accuracy profit models, it is better to combine these directly with a basic knowledge of transactions to create a reliable, applicable model~\cite{ding2015deep, hall2002predicting}.

Candlestick pattern recognition is an essential tool for determining market conditions~\cite{marshall2006candlestick}. To make trading decisions, traders often make judgments based on much-complicated information, such as technical indicators, news, and candlestick patterns. Thus, candlestick pattern recognition is a crucial support for individual transactions~\cite{bulkowski2012encyclopedia}. Candlestick pattern recognition helps traders determine the current asset price in the market and establish whether the current buying pressure will continue or whether the current selling pressure will reverse. This information, along with other sources, assists traders to predict the future. Concerning price trends, the Morning Star and the Evening Star are examples of price reversal signals commonly. Candlestick pattern recognition requires a deliberate analysis of trader expertise rather than pure numerical analysis. This recognition requires traders to make visual judgments on images.

The Convolutional Neural Network (CNN) model is well-suited to image recognition~\cite{boureau2008sparse}. CNN can update its convolution kernel by backward propagation and train the appropriate weights to extract excellent image features. The correlation between traits and images uses to help models make correct judgments. Further, the type of neural network suitable for image identification needs to carry out through a two-dimensional convolution. Principally, the financial time-series data representing uses a one-dimensional array. Therefore, we need to find a way to convert the time-series data into a consistent matrix form. 
 
However, our datasets are always dynamic, and patterns in them are changing.
Hence, we need to feature engineering to extract specific time-series features. For example, space transformation models are kinds of feature engineering. There are including Singular Value Decomposition (SVD), distance metric learning, Nystr{\"o}m methods, and Distance Metric Learning (DML) approach~\cite{li2020improving}. The process of Singular Value Decomposition (SVD) uses for investigation of the data. In these methods, linear algebra uses to construct a data matrix out of the collected data and to extract intrinsic features of that matrix. It is to separate elements that are similar between each subject and features that differentiate the items.

Instances with different labels are intertwined and often linearly inseparable. This issue brings new challenges to the CNN approach~\cite{li2017classifying, aziz2018artificial}. The CNN approach considers unsuitable for directly encoding the time-series data as image pixels~\cite{gamboa2017deep}. Hence, we need a method for transforming time series data into images.

The Gramian Angular Field (GAF) has the following advantages:
\begin{enumerate}
\item The GAF provides a way to preserve temporal dependency since time increases as the position moves from top-left to bottom-right.
\item The GAF contains temporal correlations because the Gramian Angular represents the relative correlation by superposition and difference of directions for the time interval.
\item The primary diagonal of the Gramian Angular Field matrix is the particular case.
\item The diagonal of the Gramian Angular Field matrix contains the original value and angular information.
\item From the main diagonal, we can reconstruct the time series from the high-level features learned by the deep neural network.
\end{enumerate}

Hence, we use the Gramian Angular Field (GAF) to encode the time-series data~\cite{wang2015encoding} from a one-dimensional time-series array to the two-dimensional convolutional time-series matrix. The encoding data can improve the performance of the neural network in the two-dimensional convolutional time-series significantly. When the CNN model uses the GAF encoding as input, the LeNet~\cite{lecun1995convolutional} architecture can achieve outstanding results naively.

Therefore, we design a GAF-based CNN to emulate the trader to identify candlestick pattern characteristics in an experiment. We call our approach GAF-CNN. First, we use the Geometric Brownian Motion (GBM) model to simulate a volume of price data. According to Zhiguo, we set the same parameters to set the price, and its volatility is close to the real data~\cite{he2008optimal}. Second, we choose eight candlestick patterns from The Major Candlestick Signals~\cite{MajorSignals}. These eight types of pointers are Morning Star, Bullish Engulfing, Hammer, Shooting Star, Evening Star, Bearish Engulfing, Hanging Man, and Inverted Hammer. The difference between these eight candlesticks signals is subtle and will challenge a traditional CNN model.

To improve the traditional CNN model, we use the GAF-CNN to train the GBM simulation data. Our model produces outstanding performance in the simulation data. We also use real data to verify the viability of our GAF-CNN in the real-world. We expect that GAF-CNN enables the computer to look at the candlestick patterns with as much nuance as a trader. The results show a near-92\% accuracy for the GBM simulation data. We use 2010-2017 historical data of the currency exchange rate for Euro (EUR) to US dollar (USD) to test our GAF-CNN model. The experimental results achieve a $90.70\%$ accuracy. The simulation and experimental results show that GAF-CNN is suitable for shape identification in financial trading. Although this paper uses only eight of the most classical-type indicators, various morphological extensions that can be made based on GAF-CNN are feasible, such as the W-head M-bottom. We want to establish a financial vision field through this paper making computers can recognize candlestick as a human has seen. 

The remainder of this paper organizes as follows. Section 2 provides a review of the literature, and Section 3 presents our methodology. Section 4 shows the result of our experiments. Section 5 describes the discussion of Section 4. Section 6 is the conclusion of our study, and Section 7 is the overall workflow of our experimental framework.

\section{Preliminary}
\label{S:2}
\subsection{Candlestick}
Japanese start using technical analysis to trade rice in the 17th century~\cite{wagner1994trading}. While this early version of technical analysis is different from the US version initiated by Charles Dow around 1900. Many of their guiding principles are similar. In this version, price action is more important than news and earnings. All happened information reflects in the price already. Buyers and sellers move markets based on expectations and emotions. The actual price may not reflect the underlying value. According to Steve Nison, candlestick charting first appears sometime after 1850~\cite{nison2001japanese}. Much of the credit for candlestick development and mapping goes to a legendary rice trader named Honma from the town of Sakata~\cite{tudela2008secret}. His original ideas are likely modified and refined over many years of trading, eventually resulting in the system of candlestick charting used today.

\begin{figure}[h]
\centering
\includegraphics[scale=0.7]{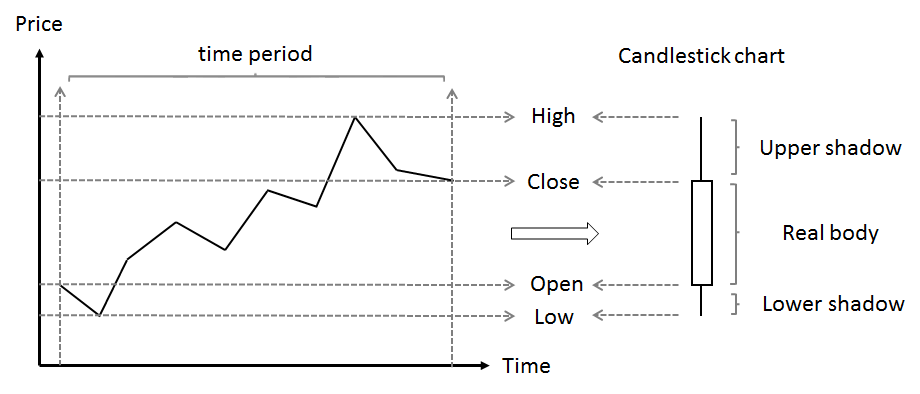}
\caption{Candlesticks display all the market needed information, such as opening, closing, high, and low prices.}
\label{candlestick}
\end{figure}
Figure~\ref{candlestick} is the structure of a candlestick. The unit is the bar, which draws on the opening, high, low, and closing prices (OHLC) for a specified period. The real-body is the price difference between the opening and closing prices. The upper shadow is the price difference between the highest price and the real-body, and the lower shadow is the price difference between the lowest price and the real-body. The period of a bar can be arbitrarily customized, usually depending on the length of the transaction. If the open price is higher than the close price, the real-body is rendered in black, indicating that the price is falling during this time. If the close price is higher than the open price, the real-body is white, indicating that the price is rising during this time. If the close price is equal to the opening price, the real-body will be just a (horizontal) line.

From the above, the candlestick helps investors filter out much of the price noise. The bar only records the different price information of OHLC per unit time. When we put together multiple bar charts, we get a continuous market information map. Unique shapes call as a pattern.

Researchers focus on the topic of candlesticks for many years~\cite{nison2001japanese}. Many patterns use to identify trends summarized, such as trend continuation indicators or reversal indicators. Candlestick analysis is an approach to getting started with trading. However, some people think it is challenging to observe the trend by observing the candlestick. It cannot use as an indicator to predict direction~\cite{goo2007application}. Human begins to systematize the patterns generated from the candlesticks. They evolve into technical indicators of the system to form the candlestick patterns gradually. The indicators are also including the Average True Range (ATR), Relative Strength Index (RSI), Moving Average (MA), Moving Average Convergence and Divergence (MACD), Stochastic Oscillator (KD)~\cite{taylor1992use} and so on.

\subsection{Convolutional Neural Networks (CNN)}
CNN models take advantage of the spatial properties of the data. According to Fukushima and Miyake, they propose a Neocognitron model. The model considers inspiring CNNs from the computational perspective generally~\cite{fukushima1982neocognitron}. Neocognitron is a neural network designed to simulate the human visual cortex~\cite{fukushima1982neocognitron}, which consists of two types of layers. The first type is the feature extractor layers, and the second type is the structured connection layers. The feature extractor layers, also named S-layers, simulate the cell in the primary visual cortex and help human beings to perform feature extraction. The structured connection layers, also named C-layers, affect the complex cell in the higher pathway of the visual cortex, provide the model with its shifted invariant property.

\begin{figure}[htbp]
\centering
\includegraphics[scale=0.5]{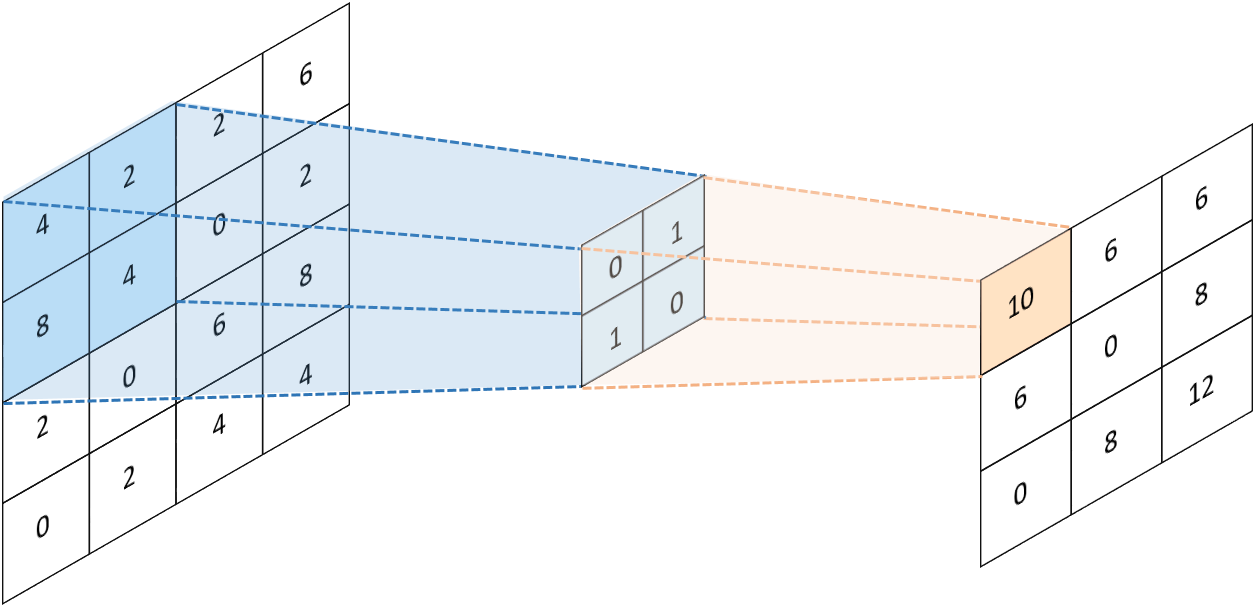}
\caption{The convolutional operation.}
\label{convolution}
\end{figure}

\begin{figure}[htbp]
\centering
\includegraphics[scale=0.5]{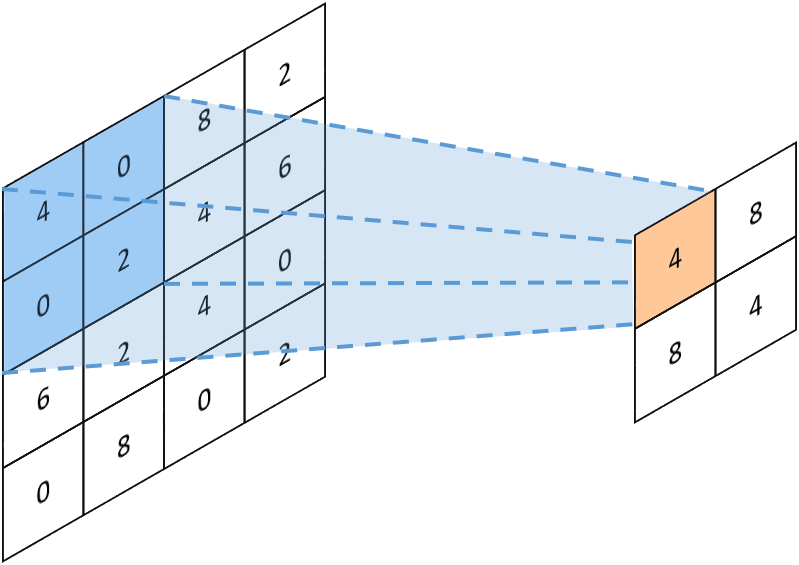}
\caption{The pooling operation.}
\label{pooling}
\end{figure}

The two most essential components of CNN are the convolutional layer and the pooling (Pool) layer. Figure~\ref{convolution} shows that the convolutional layer implements the convolutional operation, which extracts image features by computing the inner product of an input image matrix and a kernel matrix. The number of channels of the input image and kernel matrix must be the same. For example, if the input image is a red-green-blue (RGB) color space, then the depth of the kernel matrix must be three; otherwise, the kernel matrix cannot capture the information between different color spaces. The pooling layer, also called the sub-sampling layer, is mainly in charge of simplifying the task. Figure~\ref{pooling} shows that the pooling layer only retains part of the data after the convolutional layer. It reduces the number of significant features extracted by the convolutional layer and refines the remaining features.

Only with these two components can the convolutional model be used to imitate human vision. In practical applications, the CNN model usually combines the convolutional layer and the pooling layer. The convolutional layer often extracts a significant number of features, and most of the elements may be noise, which could lead to the model learning in the wrong direction, also known as over-fitting. Furthermore, the fully-connected layers connect at the end of the sequence usually. The function of the fully-connected layer organizes the extracted features processed by the convolutional and pooling layers. The correlation between the extracted features learns in this layer.

Although the pooling layer can reduce the occurrence of over-fitting after convolution, it is inappropriate to use after the fully-connected layer. The other widely recognized regularization technique, called drop-out, designs to solve this issue. The drop-out technique randomly drops neurons with a specific probability, and the dropped neurons are not involved in the forwarding and backward computation. This idea directly limits the model's learning; the model can only update its parameters subject to the remaining neurons in each epoch.

\begin{figure}[!htbp]
\centering
\includegraphics[scale=0.4]{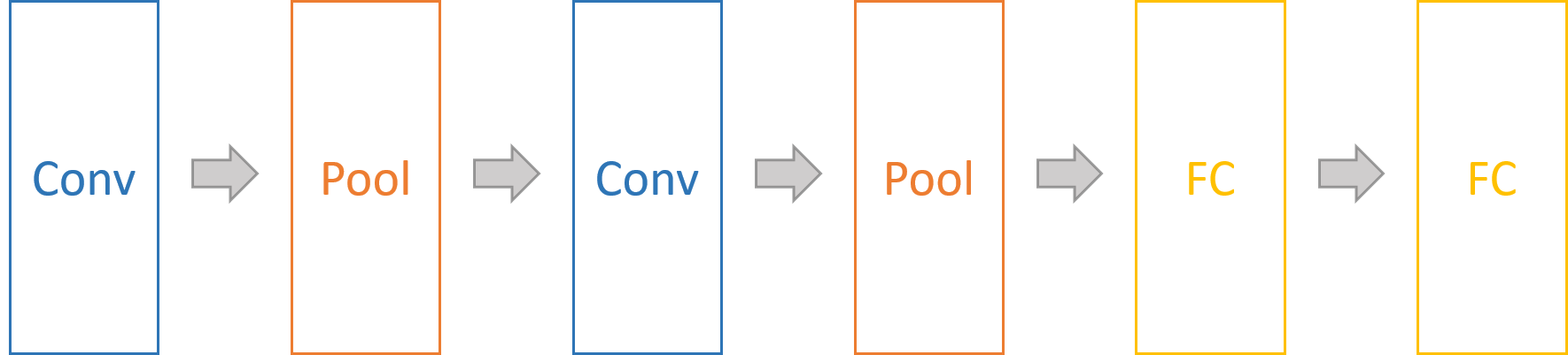}
\caption{The classic LeNet model.}
\label{lenet}
\end{figure}

The most general classic modern CNN model, LeNet inspires by Neocognitron and the concept of backpropagation~\cite{lecun1995convolutional}. The potential of the modern convolution architecture can be seen in LeNet~\cite{lecun2015lenet}, consisting of a convolution layer, a subsampling layer, and a full connection (FC) layer~\cite{wang2017origin}. Figure~\ref{lenet} shows the LeNet model. As the concept of the rectified linear unit (ReLU) and drop out are presented in recent years, a new convolution-based model, AlexNet, proposed by Alex Krizhevsky and Hinton~\cite{krizhevsky2012imagenet}, appeared and beat the previous champion of the ImageNet Challenge, with 10M labeled high-resolution images and 10,000+ object categories.

\subsection{CNN for Patterns Classification}
Human beings are visual creatures. The eyes are the most compact structure of all the sensory organs, and the visual intelligence of the human brain is rich in content. Exercise, behavior, and thinking activities all use visual sensory data as their most significant source of information. The more flexible and talented we become, the more we rely on visual intelligence. What general business and decision-makers desire after the analysis is not the data itself, but the value. Therefore, data analyses must be intuitive. In this way, the visualization of financial data more readily accept: decision-makers can see the story and interpret the data more efficiently.

Although visualization analysis can benefit decision-makers, many traditional statistical or machine learning methods for predicting currency movements use quantitative models. These methods do not consider visualization. We attempt to make good use of the advantages of display and comprehensively enhance the efficiency of intelligence analysis. For example, most traders use charts to analyze and predict currency movement trends, which carry apparent economic benefits. However, in this visualization, the analysis is artificial. We aim to teach machines to achieve the interpretation of visual information like a human brain. We then hope to use the tool to analyze robust financial data visually.

The CNN models use in pattern and image recognition problems widely. In these applications, the best possible accuracy has achieved using CNNs. For example, the CNN models have achieved a accuracy of 99.77\% using the Modified National Institute of Standards and Technology (MNIST) database of handwritten digits~\cite{ciregan2012multi}, a accuracy of 97.47\% with the New York University Object Recognition Benchmark (NORB) dataset of 3D objects, and a accuracy of 97.6\% on over 5,600 images of more than ten objects. The CNN models not only give the best performance compared to other detection algorithms but also outperform humans in such cases as classifying objects into fine-grained categories, such as particular breeds of dogs or species of bird. The two main reasons for choosing a CNN model to predict currency movements are as follows:
\begin{enumerate}
\item The CNN models are good at detecting patterns in images, such as lines. We expect that this property can use to detect trends in trading charts. 
\item The CNN models can detect relationships among images that humans cannot find easily. The structure of neural networks can help detect complicated relationships among features.
\end{enumerate}

\subsection{Gramian Angular Field (GAF)}
GAF is a novel time-series encoding method proposed by Wang and Oates~\cite{wang2015encoding}, which represents time series data in a polar coordinate system and uses various operations to convert these angles into symmetry matrix. Gramian Angular Summation Field (GASF) is a kind of GAF using the cosine function. Each element of the GASF matrix is the cosine of the summation of angles.

Our first step to making a GAF matrix is to normalize the given time series data X into values between $[0, 1]$. The following equation shows the simple linear normalization method, where notation $\widetilde{x}_{i}$ represents the normalized data.
\begin{align}
\widetilde{x}_{i}&=\frac{x_i-\min(X)}{\max(X)-\min(X)}
\end{align}
After normalization, our second step is to represent the normalized time series data in the polar coordinate system. The following two equations show how to get the angles and radius from the rescaled time series data.
\begin{align}
\phi&=\arccos(\widetilde{x}_i), -1\leq \widetilde{x}_i \leq 1, \widetilde{x}_i \in \widetilde{X}\\
r&=\frac{t_i}{N}, t_i\in\mathbb{N}
\end{align}
Finally, we sum the angles and use the cosine function to make the GASF by the following equation:
\begin{align}
\textup{GASF}=\cos(\phi_i + \phi_j) = \widetilde{X}^T \cdot \widetilde{X} - \sqrt{I-\widetilde{X}^2}^T\cdot \sqrt{I-\widetilde{X}^2}
\end{align}

The GASF has two essential properties. First, the mapping function from the normalized time series data to GASF is bijective when $\phi \in [0,\pi]$. In other words, normalize data to $[0, 1]$ can transform the GASF back into normalized time series data by the diagonal elements. Second, in contrast to Cartesian coordinates, the polar coordinates preserve absolute temporal relations.

\section{Methodology}
\label{S:3}
This section begins with the overall experiment design, then illustrates the method of label creation, GAF-CNN model, feature selection, and neural architecture searching, respectively.

\subsection{Experiment Design}
Considering real-world data lacking and complexity, it starts with simulation data to ensure GAF-CNN model work and progress feature selection and neural architecture search. Further, it will adopt in the empirical research on real-world data.

The simulation data are including the 2000 training data, 400 validation data, and 500 testing data from the Geometric Brownian Motion (GBM) model. Furthermore, we use EUR/USD 1-minute price data from January 1, 2010, to January 1, 2018, to label the real-world data, including 1000 training data, 200 validation data, and 350 testing data.

\subsection{Illustration of Label Creation}
We select eight of the most classic candlestick patterns based on a classic candlestick patterns textbook, The Major Candlesticks Signals, as our training target. The eight candlestick patterns we chose are Morning Star, Bullish Engulfing, Hammer, Shooting Star, Evening Star, Bearish Engulfing, Hanging Man, and Inverted Hammer. All of these patterns are reversal patterns, which capture whether the price is going to change. The first four patterns detect the price from downtrend to uptrend, and the last four patterns detect the opposite. We illustrate Morning Star and Evening Star as examples below.  
\begin{figure}[h]
\begin{center}
\includegraphics[scale=0.8]{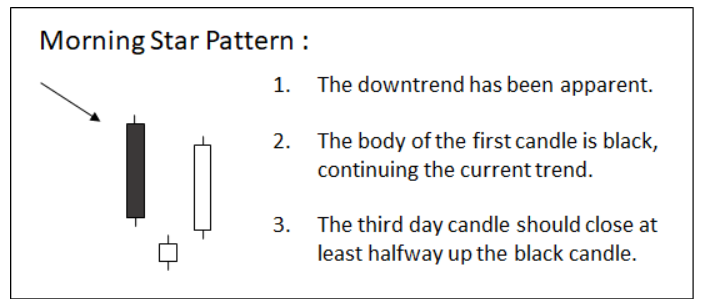}
\caption{The left-hand side shows the appearance of the Morning Star pattern. The right-hand side shows the critical rules of the Morning Star pattern.}
\label{fig:morning_star_pattern}
\end{center}
\end{figure}

The Morning Star pattern detects a price changing from a downtrend to an uptrend. The description of this pattern has three stages. First, a downtrend must be confirmed, which means the whole market has an absence of confidence. Second, the depressed atmosphere results in a big black bar. After a calm day, the third bar is a big white bar, which indicates that the investors expect the confidence of the market to reverse. Figure~\ref{fig:morning_star_pattern} shows the main appearance and rules of Morning Star in detail.

\begin{figure}[h]
\begin{center}
\includegraphics[scale=0.8]{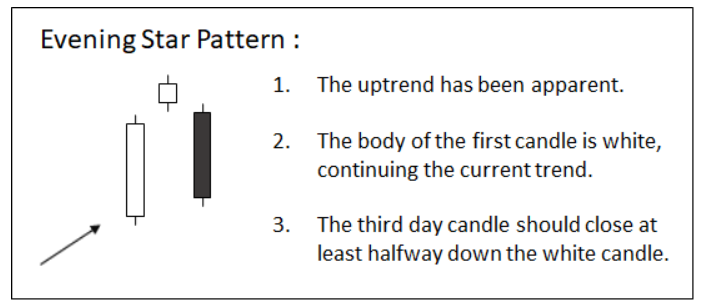}
\caption{The left-hand side shows the appearance of the Evening Star pattern. The right-hand side shows the critical rules of the Evening Star pattern.}
\label{fig:evening_star_pattern}
\end{center}
\end{figure}

\begin{figure}[h]
\begin{center}
\includegraphics[scale=0.8]{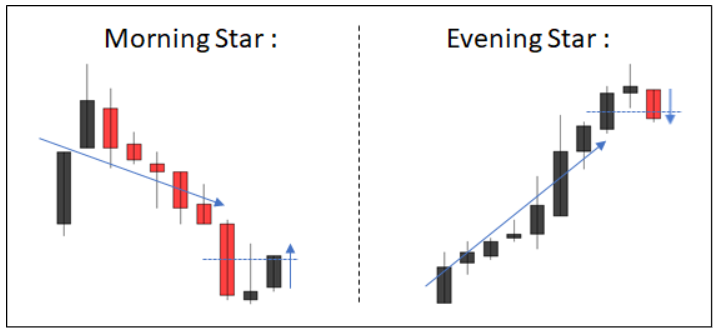}
\caption{The morning star and evening star patterns recognize in real-world data.}
\label{fig:realdata_morning_star_vs_evening_star}
\end{center}
\end{figure}

The Evening Star pattern detects the price changing from an uptrend to a downtrend. The description of this pattern also has three stages. First, an uptrend must be confirmed, which means the whole market is in a specific situation. Second, good days end with a big white bar. After a calm day, the third bar becomes a big black bar. These indicate that the investors expect the confidence of the market to reverse. Figure~\ref{fig:evening_star_pattern} shows the main appearance and rules of Evening Star in detail, and Figure~\ref{fig:realdata_morning_star_vs_evening_star} shows the difference between Morning Star and Evening Star patterns.

\begin{figure}[h]
\begin{center}
\includegraphics[scale=0.5]{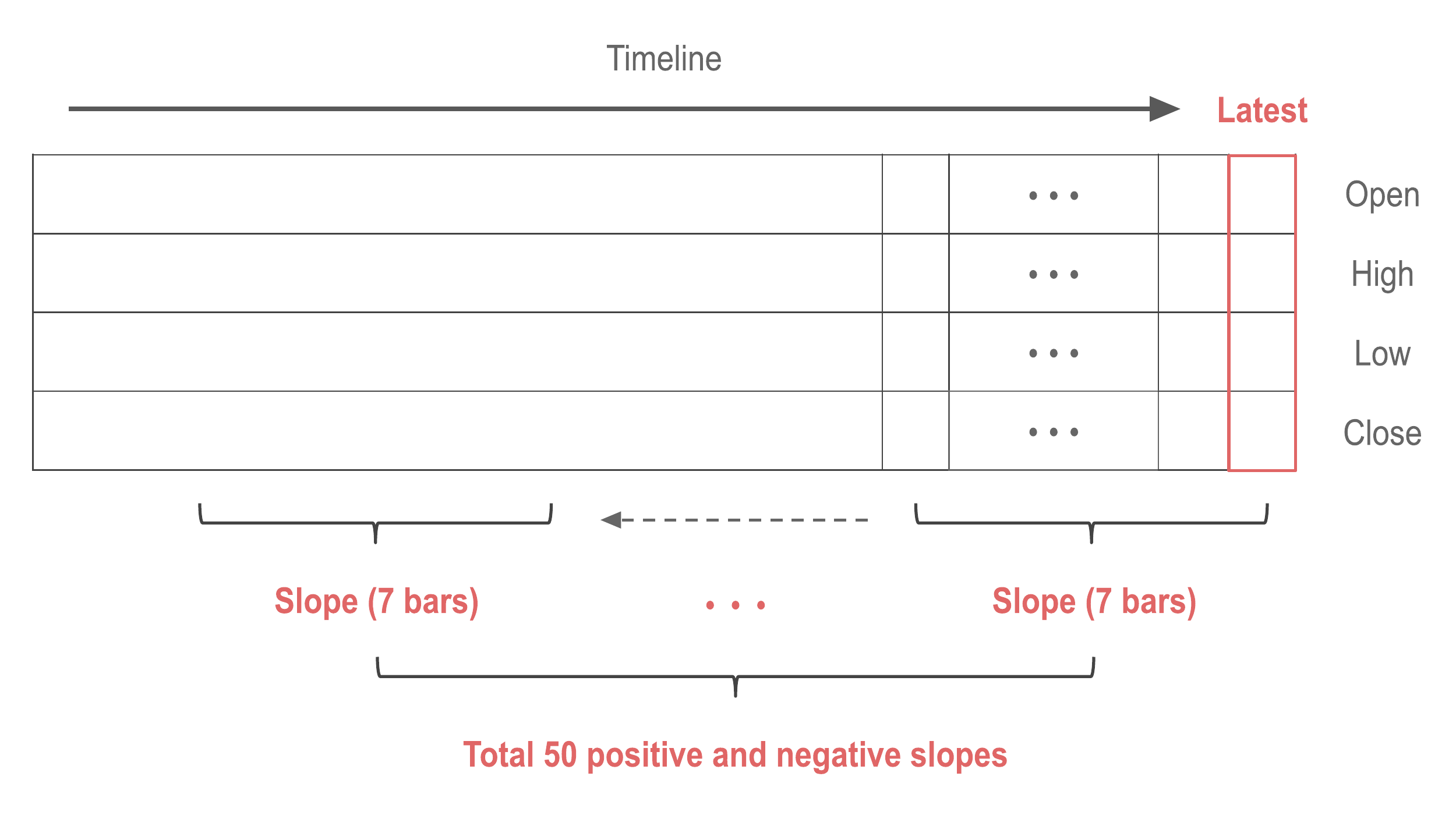}
\caption{The flowchart of our slope definition.}
\label{fig:slope}
\end{center}
\end{figure}

The definition of our label bases on the rules given in The Major Candlesticks Signals, as shown in Figures~\ref{fig:morning_star_pattern} and~\ref{fig:evening_star_pattern}. The downtrend and uptrend define from regression. If the slope is higher or lower enough, the trend is confirmed. The definition of slope in our implementation is as follows, Figure~\ref{fig:slope} has the entire illustration:
\begin{enumerate}
\item The slope value computes from the closing price among 7 bars.
\item Move a bar window to get another slope value.
\item Keep collecting positive and negative slope until 50 units, respectively.
\item If the current slope is over the 70th percentile of the group, then it will be defined as a positive or negative trend.
\end{enumerate}

We must note that the other pattern rules are slightly different between the simulation and the real data. The rules from the simulation data are similar to the book. Nevertheless, the number of samples is insufficient in real-world data because of the strictness of the rules. Hence, we relax the rules to obtain sufficient data slightly. For example, the Bullish Engulfing pattern requires the opening price of the last bar to be lower than the closing price of the previous bar. If this rule is too strict, we relax the condition such that the opening price of the last bar only needs to be less than or equal to half of the real body of the previous bar.

\subsection{GAF-CNN}
We propose a two-step approach and call it the GAF-CNN model. The first set is the Gramian Angular Summation Field (GASF) time-series encoding, and the second step is the Convolutional Neural Networks (CNN) model. In the first step, we encode time series data based on opening, high, low, and closing prices (OHLC) to GASF matrices with the window size set to $10$. After this step, the shape of the data matrices will be $(10, 10, 4)$. In the second step, we train this 3-d matrices data with the CNN model. The architecture of our second step's CNN model is similar to LeNet, including two convolutional layers with $16$ kernels and one fully-connected layer with $128$ dense. Figure~\ref{fig:methods} illustrates the entire experimental architecture, and Table~\ref{tab:gafcnn_parameters} shows the parameters used in our GAF-CNN model.

\begin{table}[]
\centering
\begin{tabular}{|c|c|}
\hline
{\sc Parameters} & \sc {\sc Values}  \\ \hline\hline
epochs           & $300$               \\ \hline
batch size       & $64$                \\ \hline
optimizer        & Adam              \\ \hline
learning rate    & $0.001$             \\ \hline
beta 1           & $0.9$               \\ \hline
beta 2           & $0.999$             \\ \hline
early stopping   & $20$ epochs         \\ \hline
\end{tabular}
\caption{The parameters of our GAF-CNN model.}
\label{tab:gafcnn_parameters}
\end{table}

\subsection{Features Selection}
\begin{figure}[h]
\begin{center}
\includegraphics[scale=0.5]{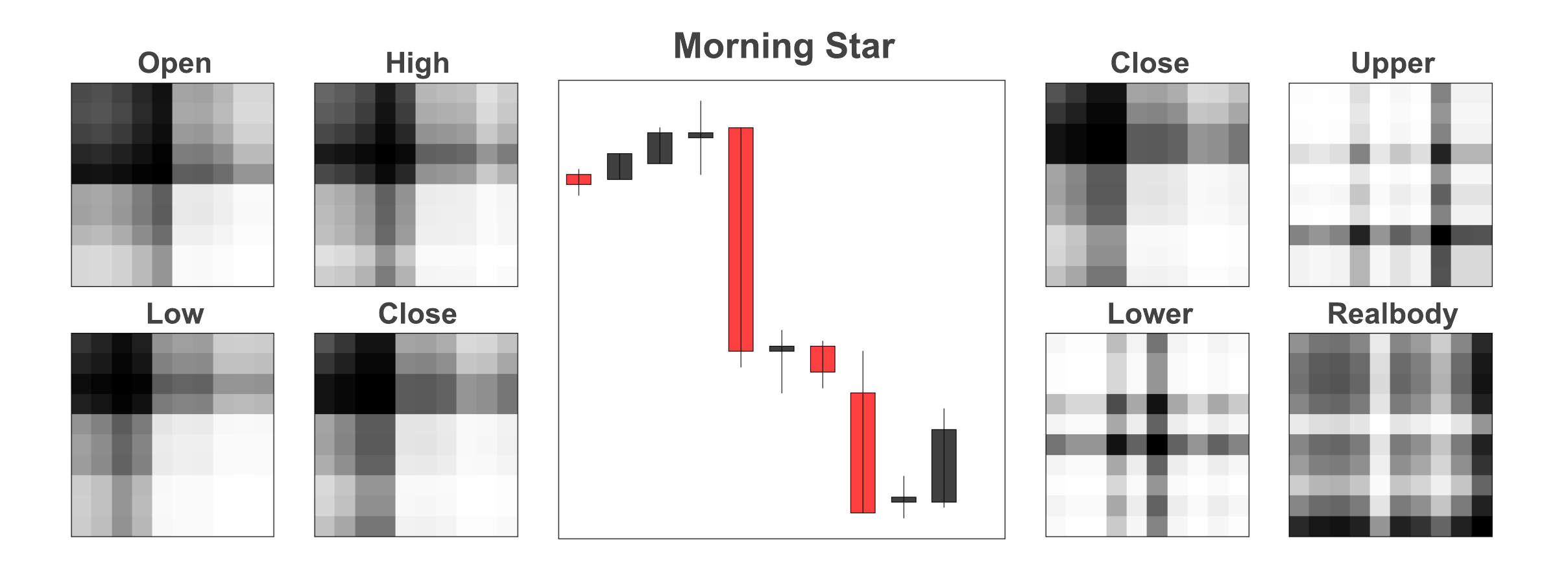}
\caption{Examples are the Morning Star patterns. The left-hand side shows the GASF features using the opening, high, low, and closing prices (OHLC) and the right-hand side shows the GASF features using closing price, upper shadow, lower shadow, and real-body (CULR).}
\label{fig:morning_ohlc_vs_clur}
\end{center}
\end{figure}

\begin{figure}[h]
\begin{center}
\includegraphics[scale=0.5]{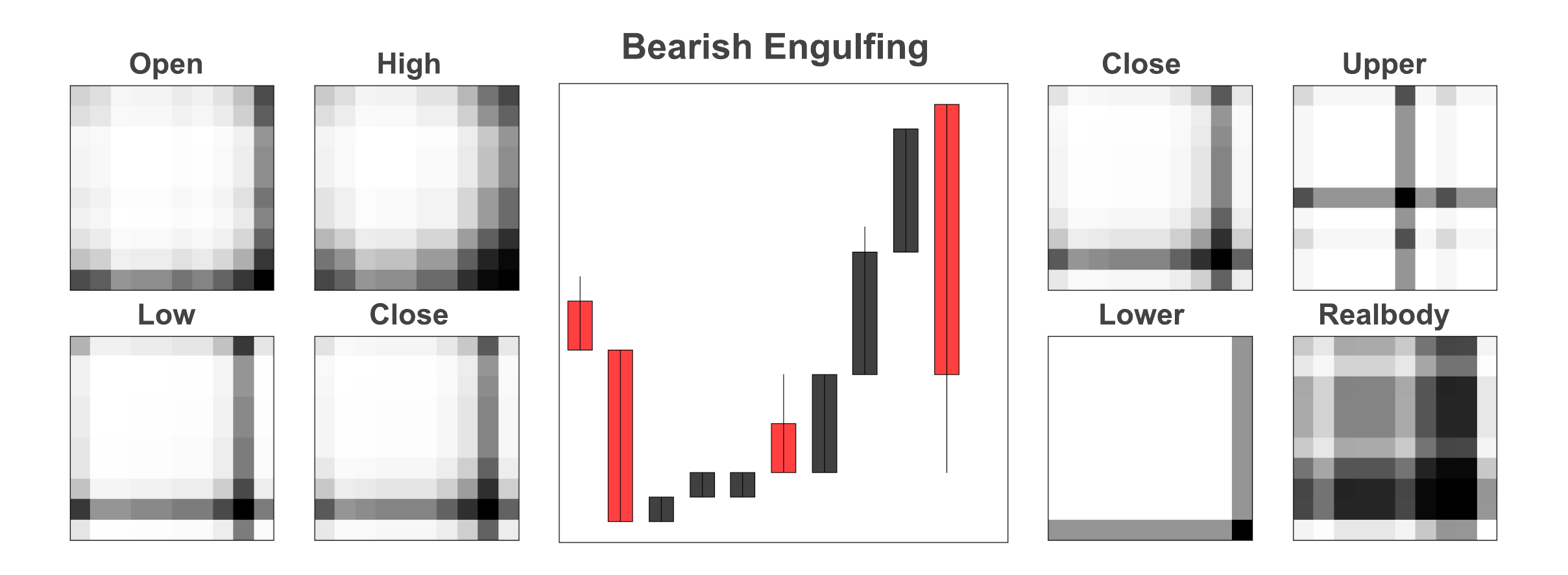}
\caption{Examples are the Bearish Engulfing patterns. The left-hand side shows the GASF features using the opening, high, low, and closing prices (OHLC) and the right-hand side shows the GASF features using closing price, upper shadow, lower shadow, and real-body (CULR).}
\label{fig:bearish_engulfing_ohlc_vs_clur}
\end{center}
\end{figure}

According to the previous section, the candlestick patterns cannot judge from a single value such as closing or opening price. Therefore, we need to combine opening, high, low, and closing prices (OHLC) and make the data features more reasonable. In order to close to humans have seen, we consider using the upper shadow, lower shadow, and real-body, which are more intuitive features for humans. Figures~\ref{fig:morning_ohlc_vs_clur} and~\ref{fig:bearish_engulfing_ohlc_vs_clur} are based on different features respectively of the Morning Star and Bearish Engulfing patterns through
\begin{enumerate}
\item the opening, high, low, and closing prices (OHLC); and
\item the closing price, upper shadow, lower shadow, and real-body (CULR).
\end{enumerate}

Figures~\ref{fig:morning_ohlc_vs_clur} and~\ref{fig:bearish_engulfing_ohlc_vs_clur} show the visualization of the GASF matrix in two kinds transformation rules. Figure~\ref{fig:bearish_engulfing_ohlc_vs_clur} shows more capable of extracting distinctive features observed than figure~\ref{fig:morning_ohlc_vs_clur}. Because the differences between the opening, high, low, and closing prices (OHLC) are generally small, resulting in high similarity among these four GASF matrices. If the model has too much repetitive information, this repeat information will reduce the convolutional model’s effectiveness in learning critical features. Hence, we process the data into the features of the second transformation rule (CULR). When we use this transformation rule, the four features are not similar and pop out the significant 2-D features in the GASF matrix. From another perspective, this is a more intuitive approach that aligns with the observations of traders. Therefore, we design our experiments using
\begin{enumerate}
\item the opening, high, low, closing prices (OHLC); and
\item the closing prices, upper shadow, lower shadow, real-body (CYLR) features
\end{enumerate}
in the simulation data.
The better results are later applied to the real-world data.

\subsection{Neural Architecture Searching}
\begin{figure}[h]
\begin{center}
\includegraphics[scale=0.4]{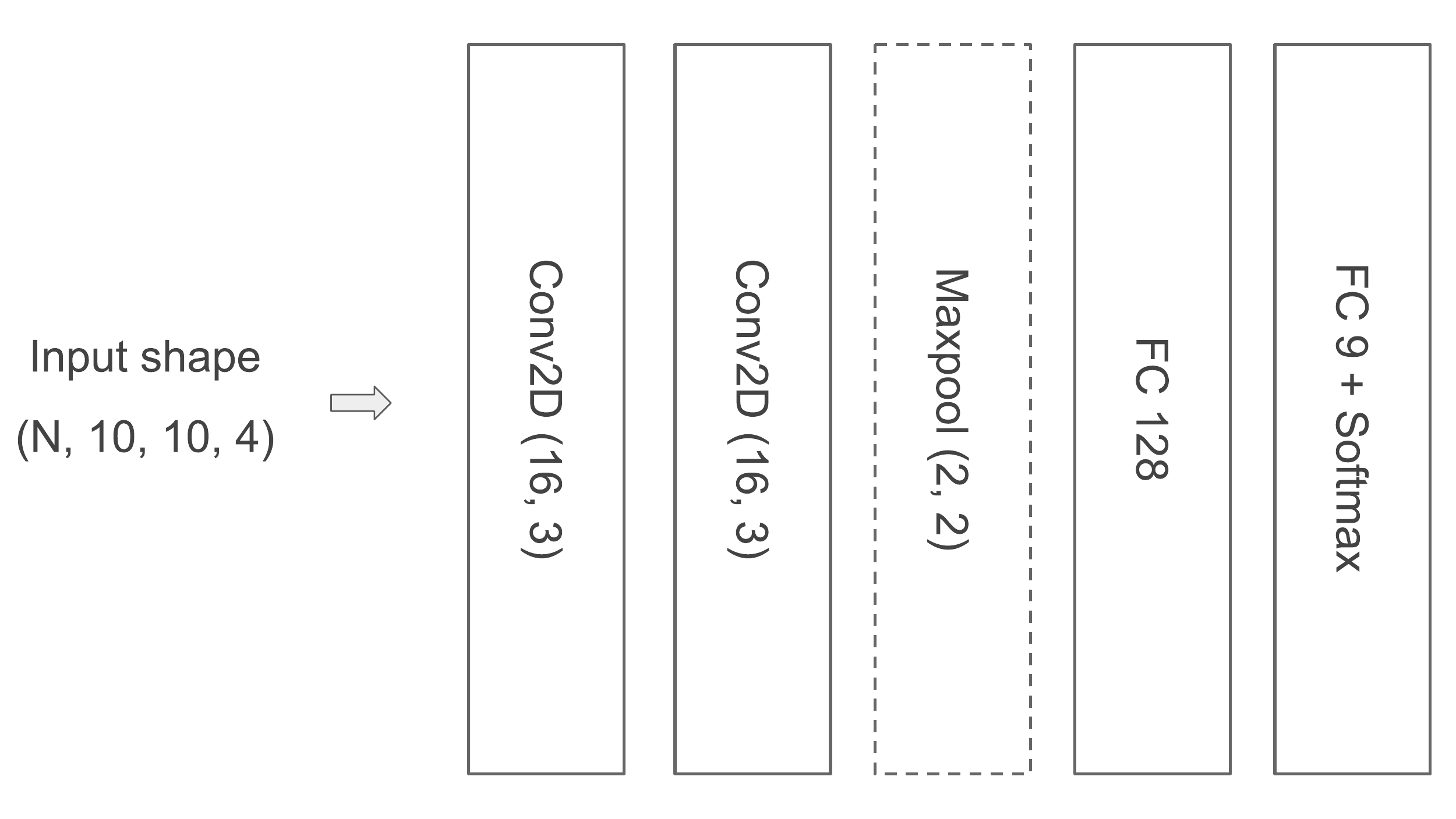}
\caption{The max-pooling use-case.}
\label{fig:pooling_or_not}
\end{center}
\end{figure}

The GAF-CNN model works well with the simple neural architecture, two convolutional layers with $16$ kernels, and one fully-connected layer with $128$ denses. The max-pooling layer, which uses general picture classification, calculates the maximum value for each patch of the feature map usually. In other words, it may bring benefits about calculating cost-saving, but truncate the characteristics of the time series, which means discard information of data. Therefore, we design an experiment using a max-pooling layer or not in simulation data. Figure~\ref{fig:pooling_or_not} illustrates where to use the max-pooling or not.

\section{Results}
\label{S:4}

\subsection{Baseline}
Previous research on the candlestick with deep learning is about trading strategy but lack of pattern classification. It is hard to find the result from other studies to compare the GAF-CNN model, so we chose the Long Short-Term Memory model (LSTM) for reliable comparison since it is a standard method to accomplish the time series classification or regression tasks in the current year. Our goal is to achieve or surpass the performance of the LSTM model. The architecture used in this study include two hidden layer size of 128 LSTM layer and follow by a 128 dense layer~\cite{smirnov2018time}. More detail comparisons will discuss in sections 4-2 and 4-3.

\subsection{Simulation Results}
\begin{figure}[!h]
\begin{center}
\includegraphics[scale=0.5]{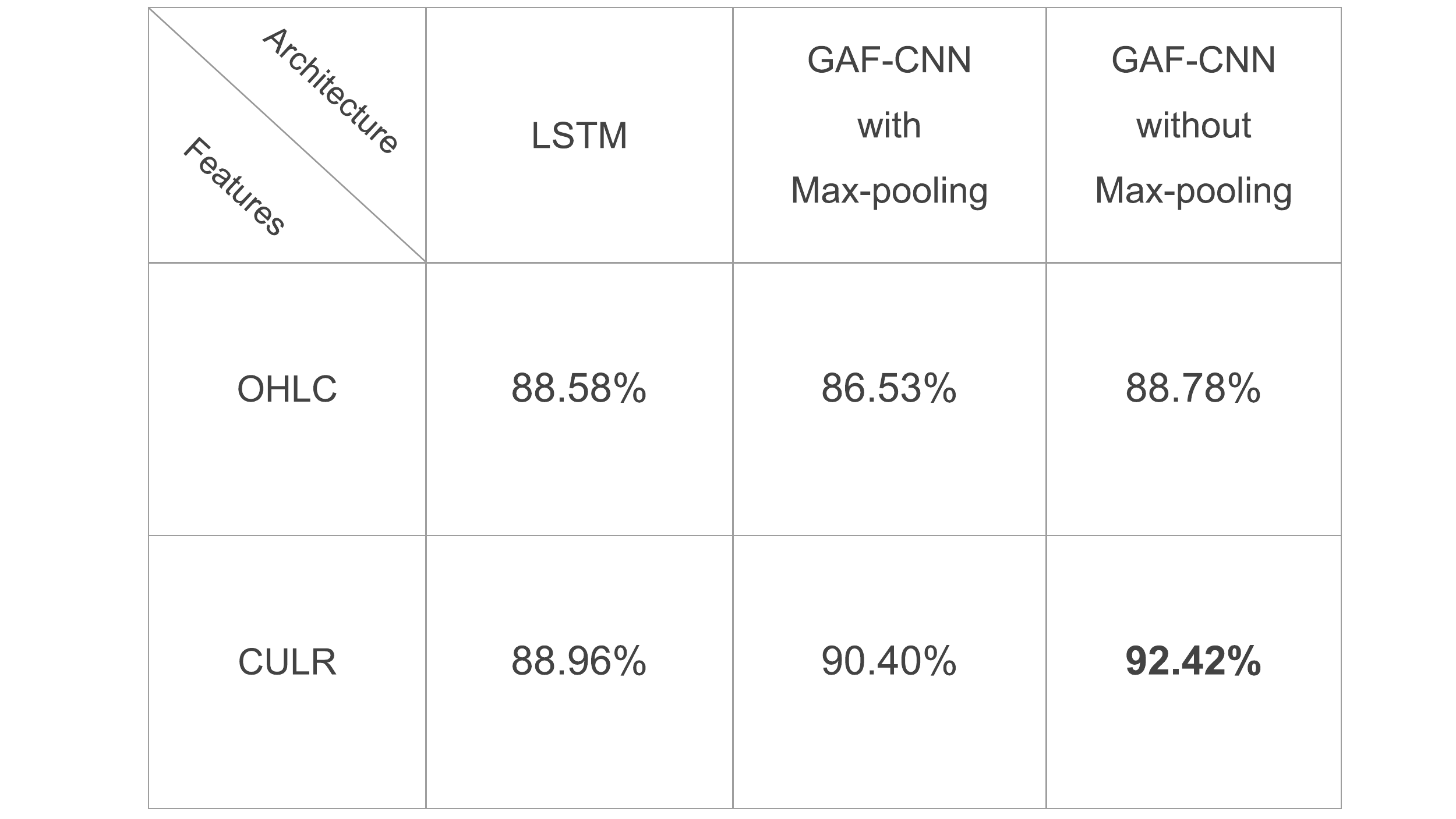}
\caption{The result is the difference between feature sets and neural architectures.}
\label{fig:compare}
\end{center}
\end{figure}

Figure~\ref{fig:compare} shows the result comparing between different features and neural architectures mention in section 3. Each experiment searches $100$ times to find out the best model and predict testing data. 

\begin{figure}[!h]
\begin{center}
\includegraphics[scale=0.5]{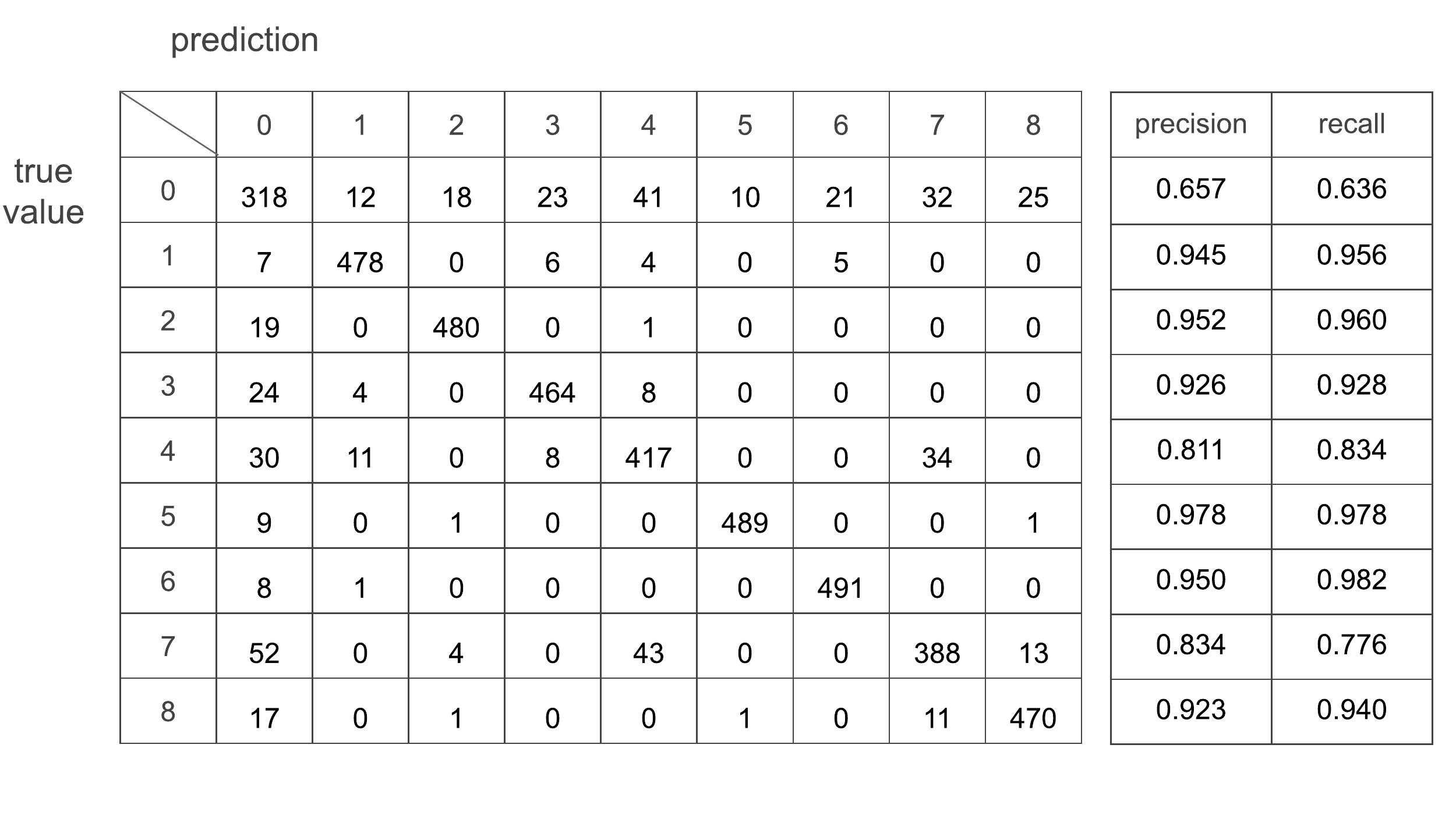}
\caption{The confusion matrix is from GAF-CNN without using pooling layers with opening, high, low, and closing prices (OHLC) feature set. The accuracy is $88.78\%$ on average.}
\label{fig:confusion_ohlc_sim}
\end{center}
\end{figure}

\begin{figure}[!h]
\begin{center}
\includegraphics[scale=0.5]{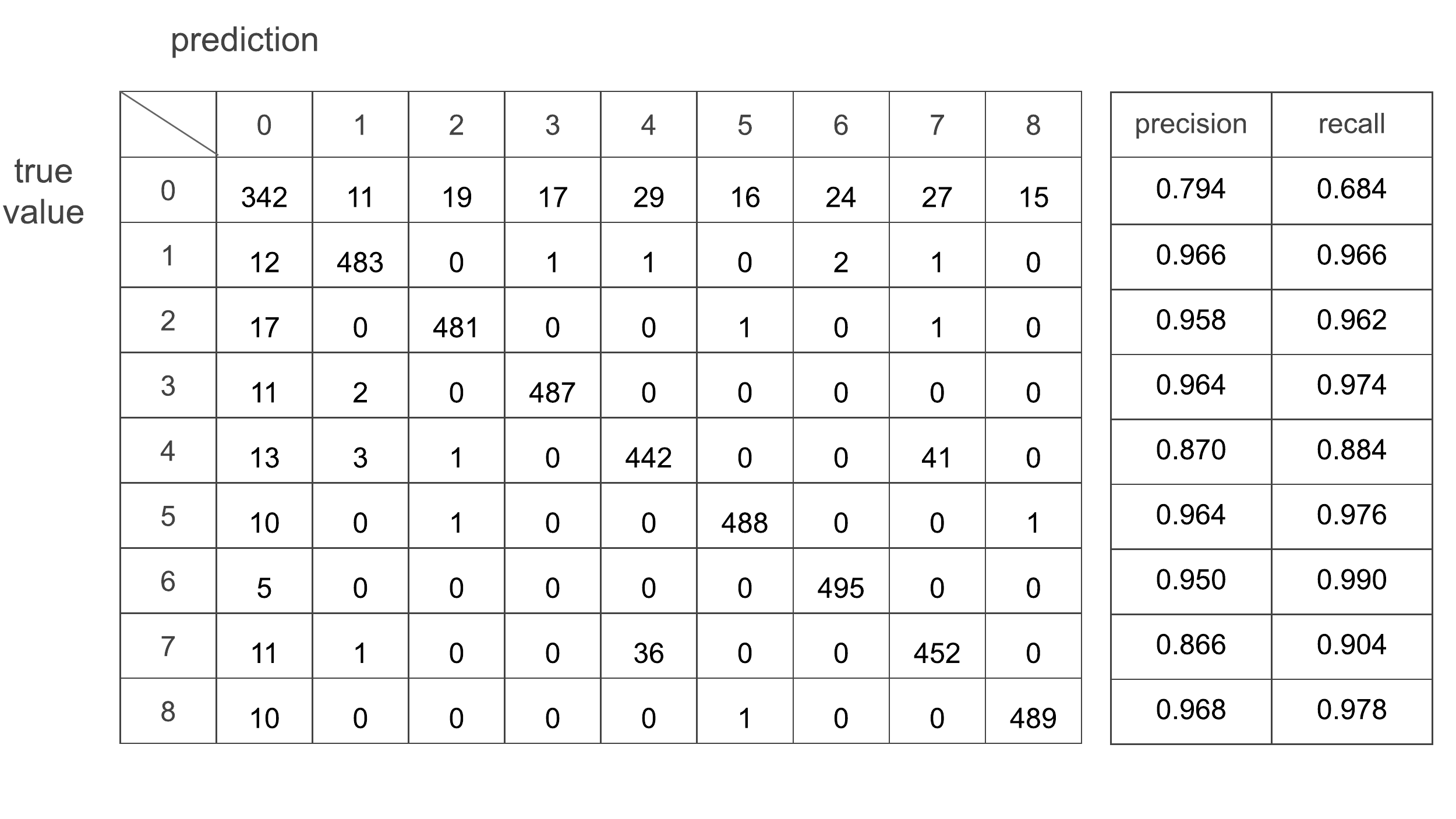}
\caption{The confusion matrix is from GAF-CNN without using pooling layers with closing, upper shadow, lower shadow, and real-body (CULR) feature set. The accuracy is $92.42\%$ on average.}
\label{fig:confusion_culr_sim}
\end{center}
\end{figure}

The GAF-CNN model without the max-pooling layer can achieve $92.42\%$ accuracy, which is better than the LSTM model $88.96\%$ accuracy in both feature sets. Figures~\ref{fig:confusion_ohlc_sim} and~\ref{fig:confusion_culr_sim} respectively show the confusion matrix of GAF-CNN model without max-pooling layer and with the different feature sets:
\begin{enumerate}
\item the opening, high, low, closing prices (OHLC); and
\item the closing prices, upper shadow, lower shadow, real-body (CULR).
\end{enumerate}
The result of using (2) closing, upper shadow, lower shadow, and real-body (CULR) can achieve $92.42\%$ average accuracy. If we focus on the result from class 1 to class 8, then the performance is $95.43\%$ accuracy on average.

\begin{figure}[!h]
\begin{center}
\includegraphics[scale=0.5]{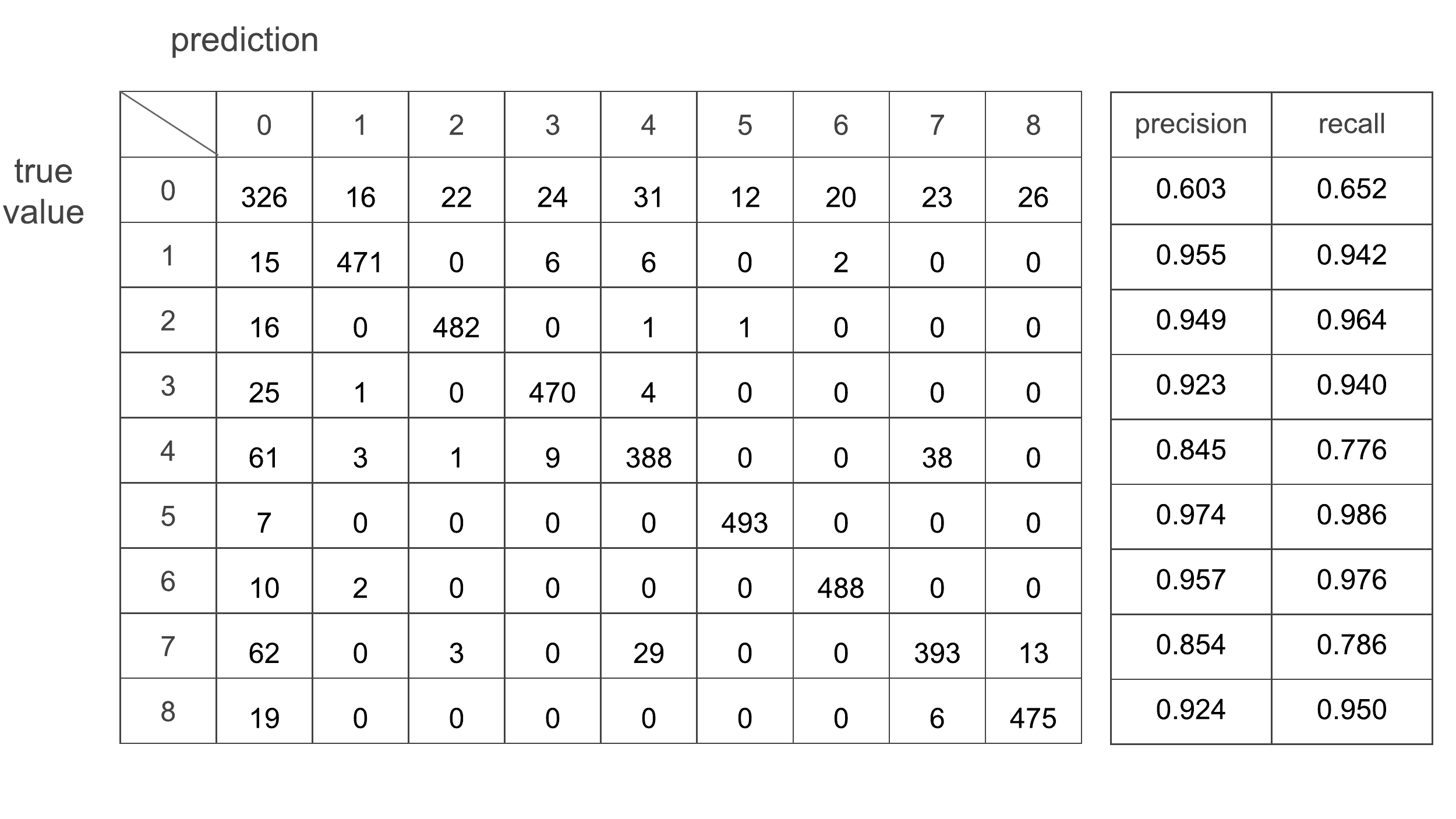}
\caption{The confusion matrix is from the LSTM model with opening, high, low, and closing prices (OHLC) feature set. The accuracy is $88.58\%$ on average.}
\label{fig:confusion_ohlc_lstm_sim}
\end{center}
\end{figure}

\begin{figure}[!h]
\begin{center}
\includegraphics[scale=0.5]{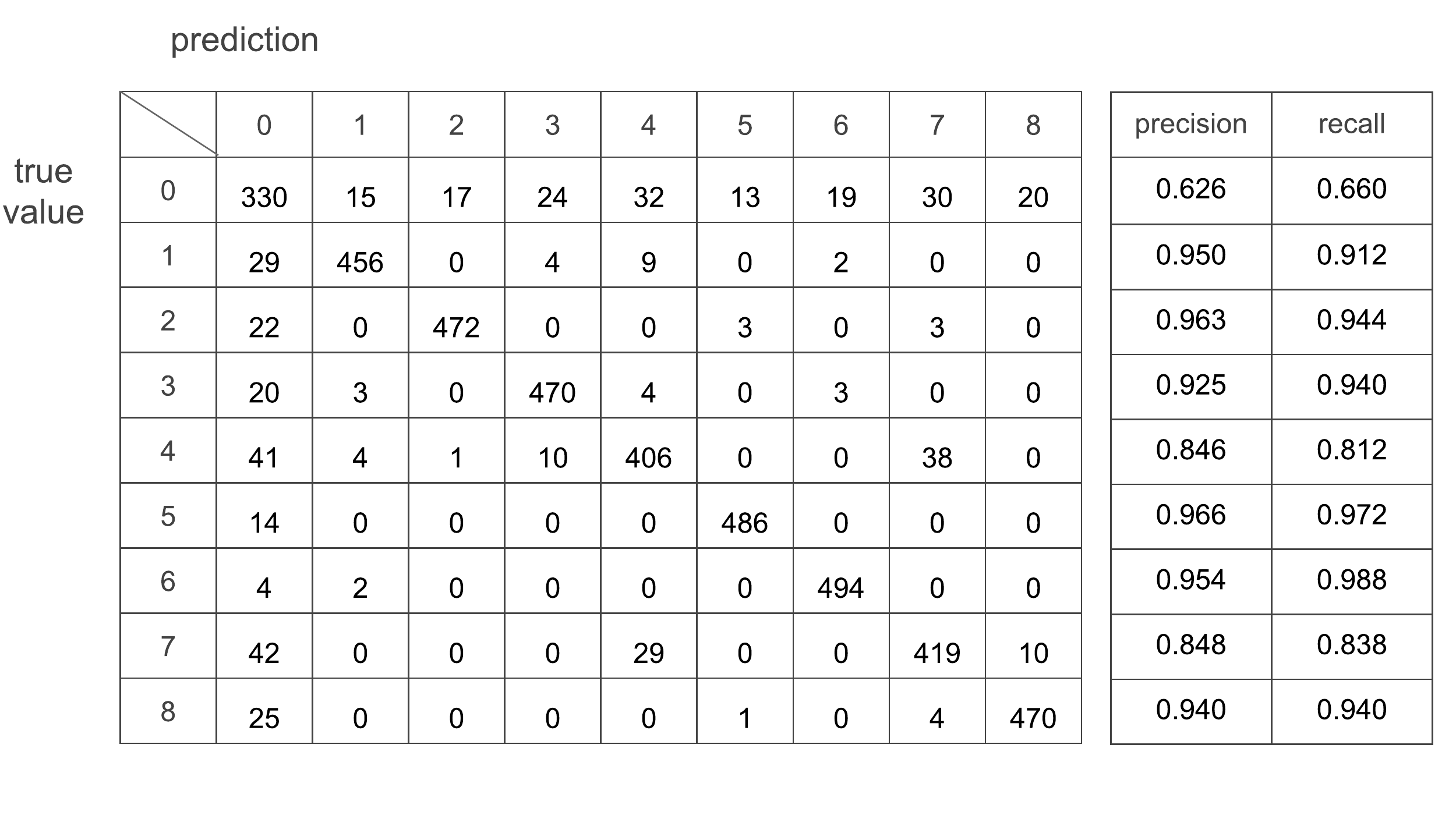}
\caption{The confusion matrix is from the LSTM model with closing, upper shadow, lower shadow, and real-body (CULR) feature set. The accuracy is $88.96\%$ on average.}
\label{fig:confusion_culr_lstm_sim}
\end{center}
\end{figure}

Figures~\ref{fig:confusion_ohlc_lstm_sim} and~\ref{fig:confusion_culr_lstm_sim} show the confusion matrix of LSTM model with two feature sets respectively. The accuracy of using (1) opening, high, low, closing prices (OHLC) is $88.58\%$ on average, and using (2) closing, upper shadow, lower shadow, real-body (CULR) is $88.96\%$ on average.

\begin{figure}[!h]
\begin{center}
\includegraphics[scale=0.21]{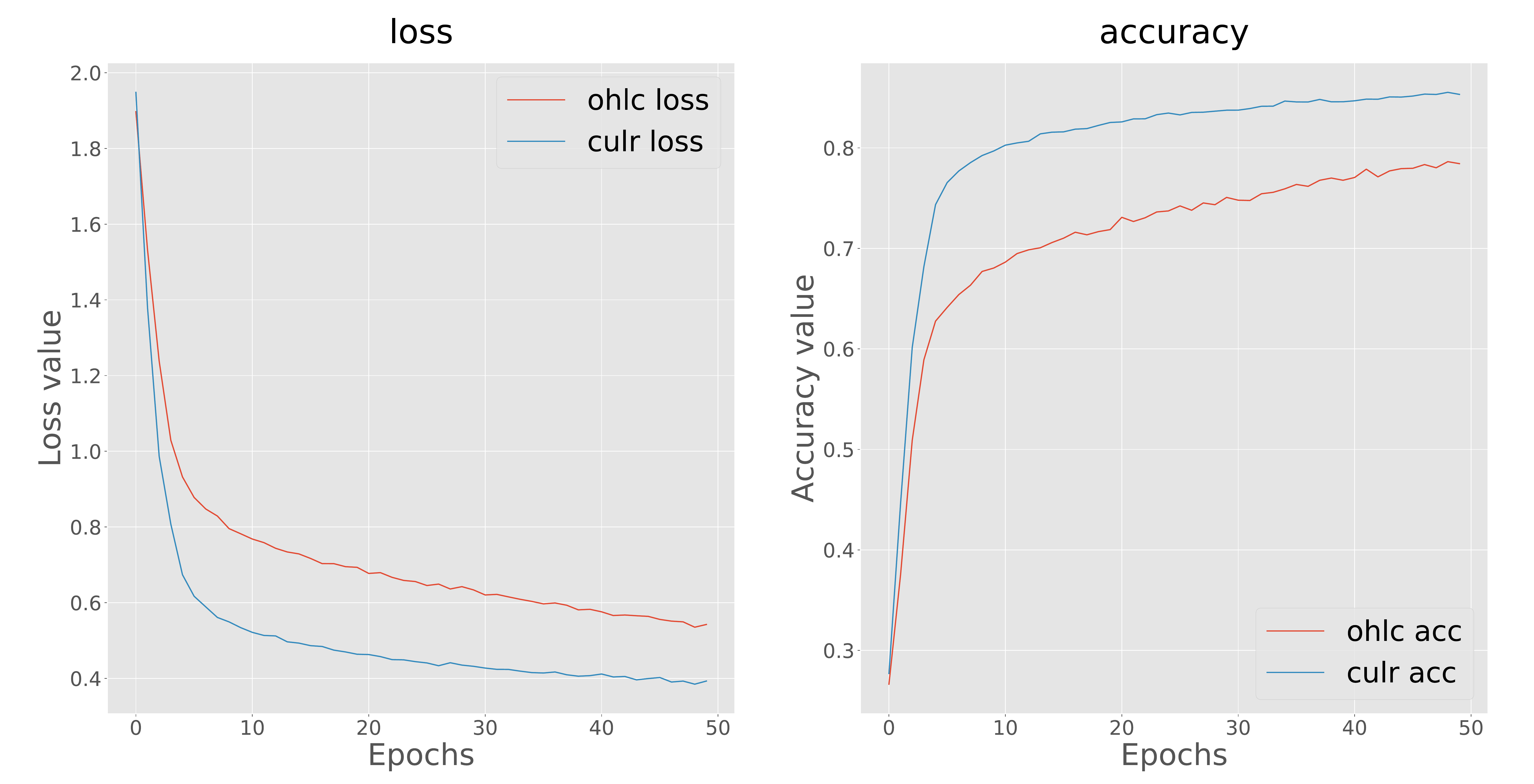}
\caption{The loss and accuracy are from the first 50 epochs within two feature sets.}
\label{fig:simulation_ohlc_vs_culr}
\end{center}
\end{figure}

\begin{figure}[!h]
\begin{center}
\includegraphics[scale=0.21]{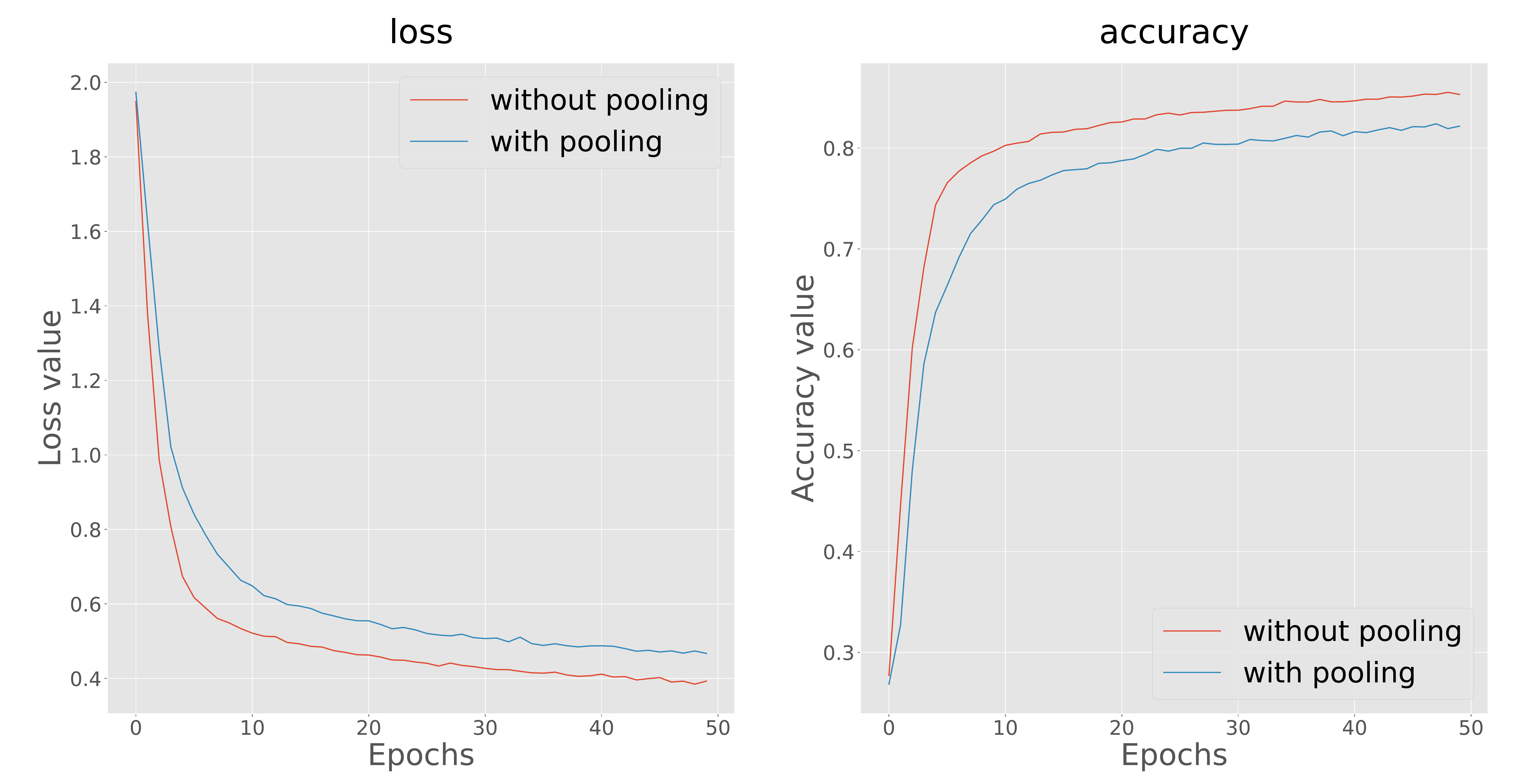}
\caption{The loss and accuracy are from the first 50 epochs between using the max-pooling layer and without using the max-pooling layer.}
\label{fig:simulation_pooling_or_not}
\end{center}
\end{figure}

To explore more about the model training process, a comparison of the first 50 epochs under different conditions would help to realize the rate of convergence. Figure~\ref{fig:simulation_ohlc_vs_culr} and Figure~\ref{fig:simulation_pooling_or_not} depict the difference of both feature sets and using max-pooling or not respectively.

\subsection{Empirical Results}
EUR/USD 1-minute price data from January 1, 2010, to January 1, 2018, are used in our real data framework, including 1000 training data, 200 validation data, and 350 testing data. Therefore, we used two times as much data in training set for class 0, which is the noisy data for the other classes. The purpose of this is to help the model clearly distinguish the patterns and increase the robustness.

\begin{figure}[!h]
\begin{center}
\includegraphics[scale=0.5]{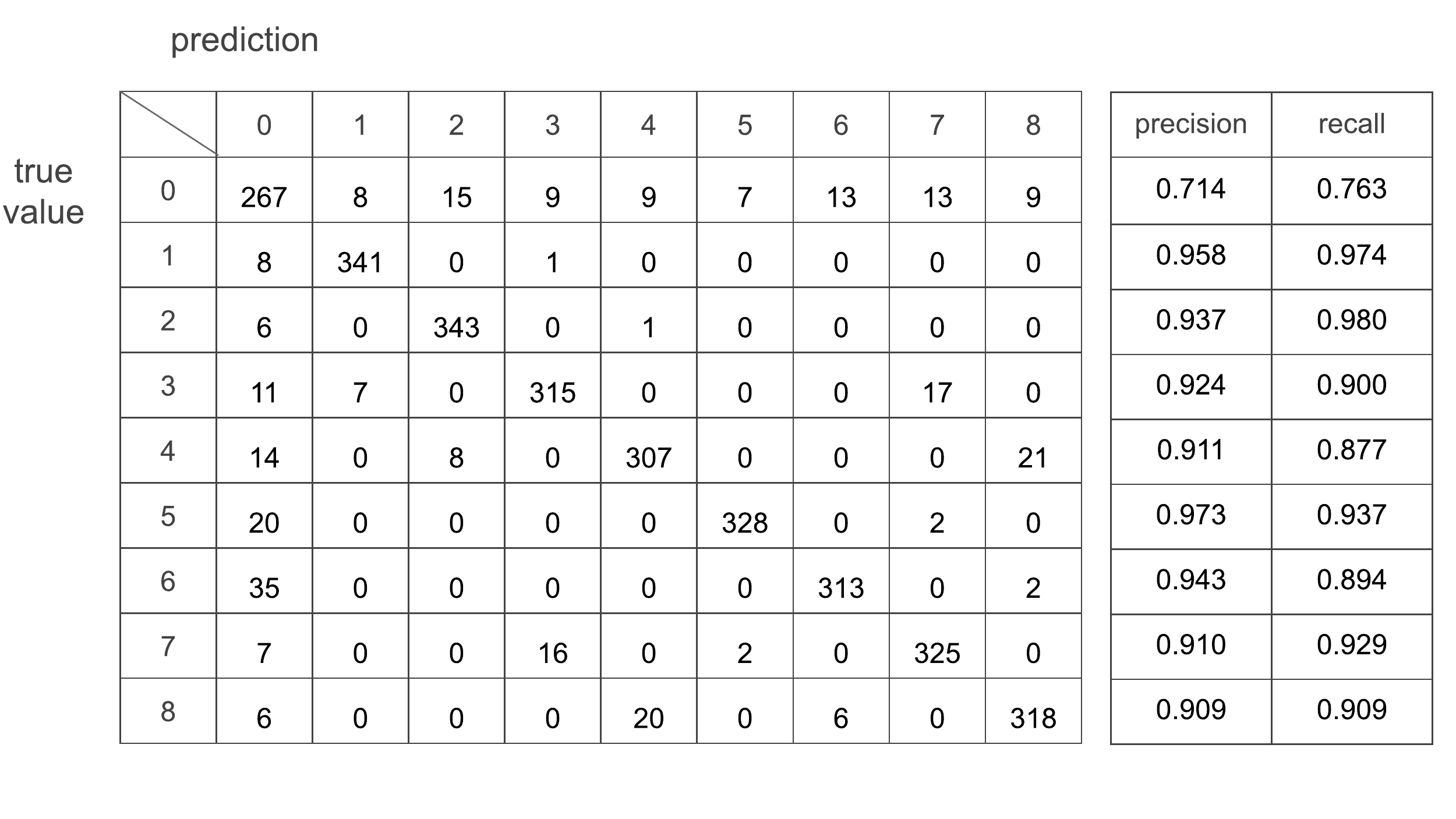}
\caption{The confusion matrix is from the real data framework. The accuracy is $90.7\%$ on average.}
\label{fig:confusion_culr_real}
\end{center}
\end{figure}

Based on the results of the simulation data, we chose to use closing, upper shadow, lower shadow, and real-body (CULR) as our feature set, and to exclude the pooling layers in our model. Figure~\ref{fig:confusion_culr_real} shows the confusion matrix of the real-world framework. The GAF-CNN model achieves $90.7\%$ accuracy on average in real-world data.

\section{Discussion}
\label{S:5}
\subsection{Simulation Results}
First of all, Figure~\ref{fig:compare} shows that using (2) closing prices, upper shadow, lower shadow, a real-body (CULR) feature set can significantly improve the accuracy in the GAF-CNN model than using (1) the opening, high, low, closing prices (OHLC) feature set. In Figure~\ref{fig:simulation_ohlc_vs_culr}, the training process also converges significantly faster in the first 50 epochs, and end up with higher accuracy. This result is intuitive that this feature set is more close to trader way, observing the characteristics of the candlestick.

Secondly, the model also converges faster when using (2) without max-pooling layer than (1) with the max-pooling layer in Figure~\ref{fig:simulation_pooling_or_not}. In Figure~\ref{fig:compare}, the GAF-CNN model without the max-pooling layer can achieve higher accuracy and lower loss value in both feature sets. The result can explain that the dependency on time series data contains many essential features. The complete time-series information will be truncated after the processing of the max-pooling layer, making it harder for the convolutional model to capture more detail features.

Lastly, the GAF-CNN model works well in both simulation and real-world framework. It achieves $90.7\%$ accuracy on average in real-world data. Besides, our results show that class 0, which is the other class, has reduced precision and recall. The class does not affect the usability of the framework because, although class 0 does not perform well, as long as the accuracy of the other classes is high enough, the cost of misclassification is small.

\subsection{Empirical Results}
The result in Figure~\ref{fig:confusion_culr_real} shows that GAF-CNN can achieve $90.7\%$ on average in the real-world data, outperforming the result of LSTM model. Therefore, our experimental results show that the GAF and the CNN framework are well-suited for candlestick pattern recognition for both simulation and real-world trading data.

\section{Conclusions}
\label{S:6}
Candlestick pattern recognition is an indicator that traders often judge with news, fundamentals, and technical indicators. However, even today, most traders decide by using their vision and experience. Although many people have directly drawn up rules to find patterns, the process is too cumbersome and hard to judge without the provision of soft scores. To better align with how traders identify patterns, we chose to use the two-dimensional CNN model. We used the GAF time series encoding with the traditional CNN model Because of the direct use of images to train leads to underfit. We use GAF-CNN to process the GBM simulation and EUR/USD real word experiments.

In the simulation framework, we use eight candlestick patterns to test how the max-pooling layer and feature sets impact our model. The results indicate the following:
\begin{enumerate}
\item The max-pooling layer is terrible for the GAF-CNN model. We think that the time series are truncated and lead to the loss of practical information.
\item Using the feature set of closing price, upper shadow, lower shadow, and real-body (CULR) is better than using the simple feature set of opening, high, low, and closing prices (OHLC).
\end{enumerate}
The model achieved an average accuracy of $92.42\%$ in simulation data. Although the 0 class is prone to misclassification, the model is still available for practical work as long as the main pattern resolutions and recall are high enough.

In the real-world framework, we use the same model for the EUR/USD per minute data from January 1, 2010, to January 1, 2018 retraining, including 1000 training data, 200 validation data, and 350 testing data. The model obtained $90.7\%$ average accuracy, outperforming the LSTM model. In real-world data, class 0 has more false positives than other types, but the main kind of recall is a certain extent. It can be considered a more conservative model. Finally, because the difference between these eight indicators is tiny, GAF-CNN has to extract subtle features. Now we only use the eight main candlestick patterns. Furthermore, future work could apply GAF-CNN to more candlestick patterns or technical indicators, such as W-head M-bottom. Thus, the entire architecture in finance candlestick, and the extensibility of the models is enormous.

\section{Workflows}
\label{S:7}
\begin{figure}[!h]
\begin{center}
\includegraphics[scale=0.5]{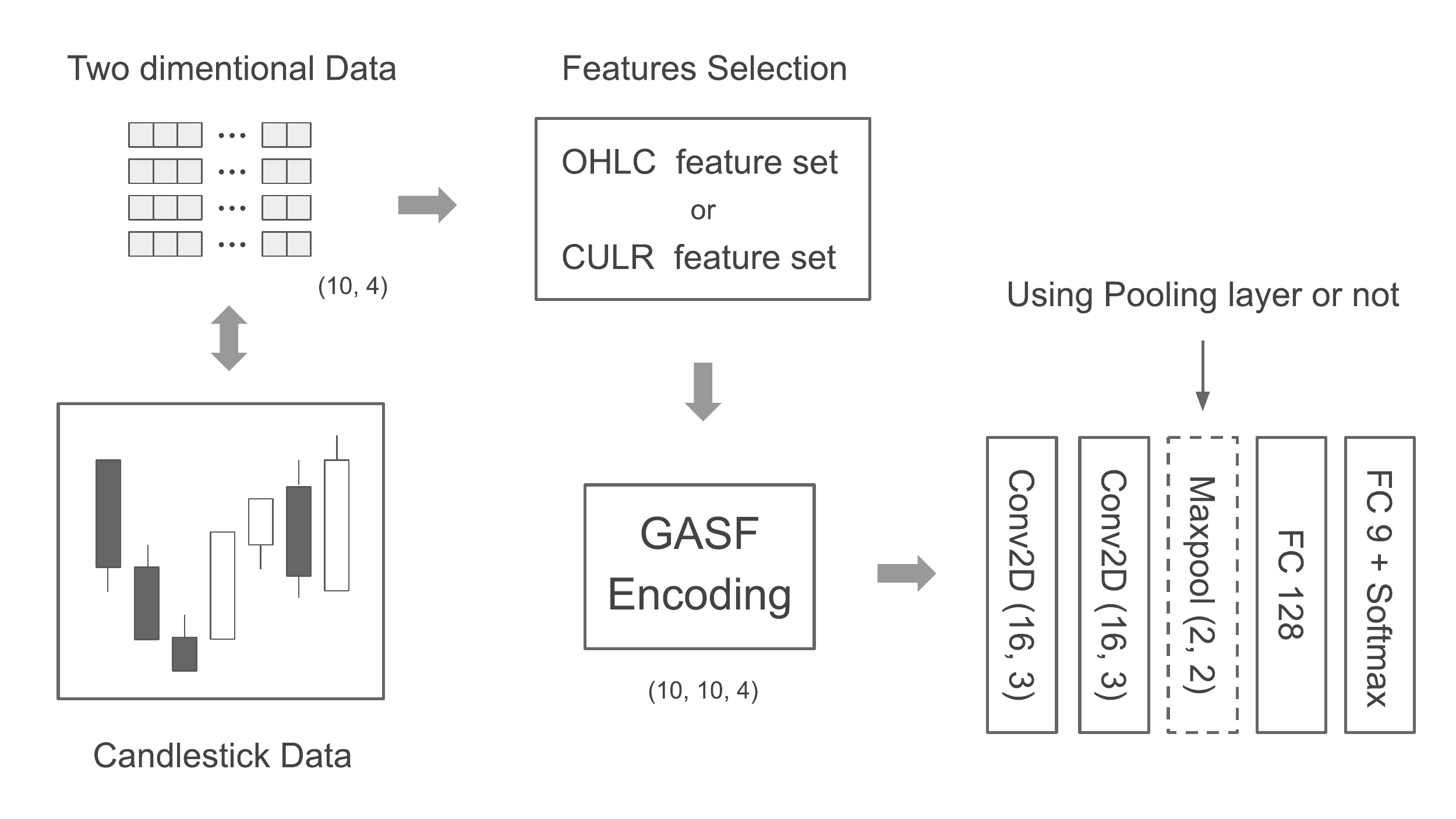}
\caption{The workflow of the entire experiment.}
\label{fig:methods}
\end{center}
\end{figure}

In this study, we find that the Convolutional Neural Network model can detect financial time series data effectively, and our research workflow is as follows:
\begin{enumerate}
\item Our experiments adopt simulation, and real-world framework, where the simulation data generates from Geometric Brownian Motion model and the real data is EUR/USD per minute data from January 1, 2010, to January 1, 2018.
\item Eight candlestick labels reference from The Major Candlestick Signals.
\item Use opening, high, low, and closing prices (OHLC) or closing, upper shadow, lower shadow, and real-body (CULR) feature sets. The data in this stage is still a 10 by 4 matrix, where 4 represents the features.
\item Encode time series data by Gramian Angular Summation Field. The data will become 10 by 10 by 4 in this stage.
\item Each framework of training, validation, and testing is with the Convolutional Neural Network model.
\end{enumerate}
The first step is each experiment test in the simulation framework, then apply the result of feature sets and neural architectures to the real-world framework. In all experiments, the convolution model use only two convolutional layers with 16 kernels and one fully-connected layer with 128 denses. All these processes illustrate in Figure~\ref{fig:methods}.

\section{Abbreviations}
\label{S:8}
\makeatletter
\newcommand{\tocfill}{\cleaders\hbox{$\m@th \mkern\@dotsep mu . \mkern\@dotsep mu$}\hfill}
\makeatother
\newcommand{\abbrlabel}[1]{\makebox[3cm][l]{\textbf{#1}\ \tocfill}}
\newenvironment{abbreviations}{\begin{list}{}{\renewcommand{\makelabel}{\abbrlabel}%
                                              \setlength{\itemsep}{0pt}}}{\end{list}}
\begin{abbreviations}
\item[ATR] Average True Range
\item[CDRs] Correction Detection Rates
\item[CNN] Convolutional Neural Network
\item[EUR] European Dollar
\item[FC] Full Connection Layer
\item[GAF] Gramian Angular Field
\item[GASF] Gramian Angular Summation Field
\item[GBM] Geometric Brownian Motion
\item[KD] Stochastic Oscillator
\item[LSTM] Long Short-Term Memory
\item[MA] Moving Average
\item[MACD] Moving Average Convergence and Divergence
\item[MNIST] Modified National Institute of Standards and Technology
\item[NORB] York University Object Recognition Benchmark
\item[OHLC] Opening, High, Low, and Closing prices
\item[CULR] Closing prices, Upper shadow, Lower shadow, Real-body
\item[Pool] Pooling Layer
\item[RSI] Relative Strength
\item[USD] United States Dollar

\end{abbreviations}

\clearpage
\section{Declarations}
\label{S:9}
\subsection{Availability of data and material}
We provide an open source (https://github.com/RainBoltz/Series2GAF) Series2GAF which can be used to transform time series into Gramian Angular Field. 
\subsection{Funding}
Jun-Hao Chen and Yun-Cheng Tsai are supported in part by the Ministry of Science and Technology of Taiwan under grant 108-2218-E-002-050-.
\subsection{Authors contributions}
Yun-Cheng Tsai conceived of the presented idea. Jun-Hao Chen developed the theory and performed the computations. Yun-Cheng Tsai and Jun-Hao Chen verified the analytical methods. All authors discussed the results and contributed to the final manuscript.
\subsection{Acknowledgements}
Thanks to Prof. Jane Yung-Jen Hsu for constructive discussion and great support.

\section{Competing interests}
Jun-Hao Chen and Yun-Cheng Tsai declare that we have no significant competing financial, professional or personal interests that might have influenced the performance or presentation of the work described in this manuscript.

\bibliographystyle{unsrt}
\bibliography{template}

\end{document}